# Physical parameters describing neuronal cargo transport by kinesin UNC-104

Kumiko Hayashi[1,2*], Shiori Matsumoto[1], Miki G. Miyamoto[1] and Shinsuke Niwa[3]

[1] Department of Applied Physics, Graduate School of Engineering, Tohoku University, Sendai, Japan

[2] JST, PRESTO, Tokyo, Japan

[3] Frontier Research Institute for Interdisciplinary Sciences (FRIS) and Graduate School of Life Sciences, Tohoku University, Sendai, Japan

**\* Corresponding author.**

Kumiko Hayashi

Department of Applied Physics, Graduate School of Engineering, Tohoku University, Sendai, Japan.

Tel:+81-22-795-7955

E-mail: kumiko@camp.apph.tohoku.ac.jp

**Abstract**

In this review, we focus on the kinesin-3 family molecular motor protein UNC-104 and its regulatory protein ARL-8. UNC-104, originally identified in *Caenorhabditis elegans* (*C. elegans*), has a primary role transporting synaptic vesicle precursors (SVPs). Although *in vitro* single-molecule experiments have been performed to primarily investigate the kinesin motor domain, these have not addressed the *in vivo* reality of the existence of regulatory proteins, such as ARL-8, that control kinesin attachment to/detachment from cargo vesicles, which is essential to the overall transport efficiency of cargo vesicles. To quantitatively understand the role of the regulatory protein, we review the *in vivo* physical parameters of UNC-104-mediated SVP transport, including force, velocity, run length and run time, derived from wild-type and *arl-8*-deletion mutant *C. elegans*. Our future aim is to facilitate the construction of a consensus physical model to connect SVP transport with pathologies related to deficient synapse construction caused by the deficient UNC-104 regulation. We hope that the physical parameters of SVP transport summarized in this review become a useful guide for the development of such model.

**Introduction**

Kinesin is a molecular motor protein that moves to the plus-end of polarized microtubules, a component of the cell cytoskeleton, by obtaining energy from adenosine-triphosphate (ATP) hydrolysis. The role of kinesin in eukaryotic cells is mainly to deliver cargo vesicles in an anterograde direction as a porter along microtubules spread throughout cells. Kinesin plays a particularly significant role in neurons, where microtubules are used as railways to convey synaptic materials from the cell center to the terminal synaptic regions via a long axon (Hirokawa et al. 2009; Vale 2003).

Physical models focusing on the motion of the kinesin motor domain and how force is generated as a result of ATP hydrolysis have been suggested (Hancock 2016; Hancock and Howard 1999; Kanada and Sasaki 2013; Peskin and Oster 1995; Sasaki et al. 2018) based on the results of *in vitro* single-molecule experiments (Nishiyama et al. 2002; Okada et al. 2003; Schnitzer et al. 2000; Tomishige et al. 2002; Vale et al. 1996; Visscher et al. 1999). However, these models do not completely explain the phenomenon of *in vivo* cargo transport, for which physical evaluations, such as force measurements, remain under development due to the complexity of the intracellular environment. Because the roles of kinesins in neurons are important to many neuronal processes (Hirokawa et al. 2009), and kinesin deficiency is related to neuronal diseases, such as Alzheimer's disease and Parkinson's disease (Chiba et al. 2014; Encalada and Goldstein 2014), physical models of *in vivo* cargo transport need to be studied in order to quantitatively understand the molecular basis of neuronal diseases. To this end, *in vivo* measurement of physical parameters associated with cargo vesicle transport needs to be summarized.

A critical difference between *in vivo* and *in vitro* analysis of kinesin-mediated transport is the existence of regulatory proteins that control cargo transport (Fig. 1a). Because these regulatory proteins control the attachment and detachment of kinesin motors with cargo vesicles, a key physical parameter describing *in vivo* cargo transport is the "number" of kinesins cooperatively carrying a single cargo vesicle (Fig. 1b). Notably, maintaining an appropriate number of motors per cargo vesicle is important for healthy neuronal activity.

Recent studies proposed a measurement method to quantify the number of motors per cargo unit *in vivo* using the fluctuation unit ($\chi$) (Hasegawa et al. 2019; Hayashi 2018; Hayashi et al. 2018a; Hayashi et al. 2018b). The fluctuation unit ($\chi$), inspired by the fluctuation theorem of non-equilibrium statistical mechanics (Ciliberto et al. 2010; Evans et al. 1993; Seifert 2012), was used to quantify force applied to cargo vesicles. Then, because the motor protein generates force acting on a cargo, the number of motor proteins can be estimated from force values. The results of these experiments revealed that several motor proteins cooperatively carry a single cargo vesicle, with measurement of fluctuation units ($\chi$) as a force indicator enabling investigation of cargo transport in relation to the number of involved motor proteins.

In this review, we summarize the physical parameters, including the number of motors, derived from previous studies (Hayashi et al. 2018a; Niwa et al. 2016) in order to evaluate regulatory systems involving *in vivo* cargo transport. Specifically, we focus on synaptic vesicle precursor (SVP) transport in the motor neurons of *Caenorhabditis elegans* (*C. elegans*), which enables green fluorescent protein (GFP)-tagged SVPs to be visualized in the living organism by fluorescence microscopy. Therefore, this represents a true *in vivo* cargo-transport system appropriate for the measurement of kinesin-related characteristics in a complex biological environment. Additionally, this system allows evaluation of SVP anterograde transport by one kinesin (UNC-104), whereas mammalian neurons use several kinesins to transport SVPs.

In the following sections, we introduce UNC-104 and its associated regulatory mechanisms in *C. elegans* motor neurons. We then present the results of physical measurements involving UNC-104-mediated SVP transport and compare them between wild-type (WT) and mutant *C. elegans* lacking a primary regulatory protein responsible for UNC-104 activity. Furthermore, we describe synapse mislocation caused by deficient UNC-104 regulation and present future perspective, which focus on the development of a physical model to potentially explain synapse mislocation as a function of the basic physical parameters involved in UNC-104-mediated SVP transport.

**UNC-104 function and regulation**

Here, we describe UNC-104 function involving SVP transport along microtubules, as well as its autoinhibition, regulated by ARL-8. Additionally, we describe fluorescence observation of UNC-104-mediated SVP transport in axons to allow analysis of UNC-104 motion.

*UNC-104 characteristics*

UNC-104 is a member of the kinesin-3 family of motor proteins originally identified by genetic analysis of *C. elegans*. UNC-104-mediated SVP transport was determined in *unc-104* mutant organisms, in which accumulation of synaptic vesicles was observed within cell bodies accompanied by dysfunctional synapse localization (Hall and Hedgecock 1991; Otsuka et al. 1991). Previous studies show that the UNC-104 homologs KIF1A and KIF1Bβ transport SVPs in mammalian neurons (Niwa et al. 2008; Okada et al. 1995; Zhou et al. 2001), and that their tail domains bind to cargos when they transport SVPs (Klopfenstein and Vale 2004; Niwa et al. 2008) (Fig. 2a and 2b). Although it was proposed that UNC-104/KIF1A could undertake processive movement as a monomer according to *in vitro* single-molecule experiments (Okada et al. 2003), the protein is considered to function as a dimer in *in vivo* (Tomishige et al. 2002).

*UNC-104 autoinhibition*

Several studies demonstrated UNC-104 inactivation via an autoinhibitory mechanism thought to be required to avoid unnecessary cargo transport, as well as avoid unnecessary ATP hydrolysis in neurons, in situations where kinesin motors do not bind cargo (Verhey and Hammond 2009). UNC-104 contains four coiled-coil (CC) domains (Fig. 2b) (Hammond et al. 2009), with the neck CC domain (NC) essential for formation of a stable dimer. The CC1 and CC2 domains are thought to negatively regulate the activity of the motor domain, and the CC3 domain binds the small GTPase ARL-8 (Niwa et al. 2016; Wu et al. 2013). A recent structural study shows that the CC1 domain directly binds to the motor and NC domains in order to inhibit dimerization and motility; however, it remains unknown how the CC2 domain negatively regulates motor activity.

A recent study described UNC-104 autoinhibition (Niwa et al. 2016) (Fig. 3). ARL-8, an SVP-bound arf-like small GTPase related to UNC-104, is essential for axonal transport of SVPs (Klassen et al. 2010) and directly binds to the UNC-104 CC3 domain in order to negate its autoinhibition (Niwa et al. 2016; Wu et al. 2013). Notably, ARL-8 is not required for UNC-104 binding to SVPs according to autoinhibition-defective UNC-104 mutants, which retain vesicle-binding ability in the absence of ARL-8 (Niwa et al. 2016). Previous reports (Niwa et al. 2008; Wagner et al. 2009) suggest that cargo-vesicle-bound ARL-8 binds to UNC-104 and releases autoinhibition by enabling binding of the tail domain to the cargo vesicle. Consequently, cargo binding induces dimerization (Klopfenstein et al. 2002), leading to full activation of UNC-104 (Tomishige et al. 2002).

*Fluorescence observation of synapses and UNC-104-mediated SVP transport*

SVPs in DA9 motor neurons of *C. elegans* can be observed by fluorescence microscopy. The *C. elegans* worms are anesthetized and fixed between cover glasses with highly viscous media prior to imaging in order to minimize the noise generated by fluctuations associated with their movement (Fig. 4a) (Hayashi et al. 2018a; Niwa 2017; Niwa et al. 2016). SVPs were labelled with GFPs (Fig. 2a) (Niwa et al. 2016). Because many SVPs accumulate at synapses, where they are unloaded from UNC-104 motors, this enables identification of synapse location by strong bright spots proximal to axonal terminal regions while SVPs can be identified as weak spots moving along the axon (Fig. 4b).

**Assumptions applied to the constant velocity segment (CVS)**

In this section, we summarize the assumptions and current model used to analyse the time courses of SVPs obtained by the fluorescence observation (Fig. 4) (Hayashi et al. 2018a; Hayashi et al. 2018b). Note that time intervals exist, during which a single SVP can be tracked (Fig. 5a, red arrow), despite axons being surrounded by numerous SVPs (Fig. 5a). Then, recorded images of fluorescence observations allow acquisition of the center position ($X$) of an SVP along an axon as a function of time ($t$) for anterograde transport (Fig. 5b).

Sharp changes in velocity are often observed in studies tracking cargo vesicles over longer time courses (Fig. 5c and 5d) and typical of *in vivo* cargo transport. In this review, such time-dependent changes in velocity are mainly considered as a result of changes in the number of motors carrying each individual cargo (Fig. 6a). Notably, this mechanism is thought to act through the repeated stochastic attachment and detachment of motors from microtubules. Then, velocity change for a given cargo can be explained by a change in the number of motors based on the force-velocity relationship of kinesin (Fig. 6b), when the viscosity effect is high enough. This suggests that long time courses should be divided into several constant velocity segments (CVSs) when the number of motors transporting a cargo is an object to be measured (Hasegawa et al. 2019; Hayashi 2018; Hayashi et al. 2018a; Hayashi et al. 2018b).

Finally, we impose one more important assumption on CVS that there is an absence of competition (tug-of-war) between kinesin and dynein (Gross 2004; Muller et al. 2008; Welte 2004), where dynein represents a molecular motor protein that moves toward the minus-end of microtubules (Vale 2003). This is supported by a recent report suggesting that kinesins do not undergo a tug-of-war with dynein during their movement in opposite directions on the microtubule (Serra-Marques et al. 2019). It is an important future issue to further study this assumption.

**Measurement of the physical parameters associated with SVP transport**

This section describes the physical parameters involved in SVP transport measured in the previous reports (Hayashi et al. 2018a; Niwa et al. 2016), and their comparison between WT and *arl-8*-deletion mutant *C. elegans* (Niwa et al. 2016).

*Force*

In Fig. 5b, the SVP showed the directional motion transported by UNC-104 while exhibiting fluctuating behaviour originated mainly from thermal noise, stepping of the motors, and collisions of the SVP with other vesicles and cytoskeletons. We consider to quantify force acting on the SVP from its fluctuating behaviour.

For each CVS (Fig. 5b), the fluctuation unit ($\chi_{FT}$) is defined as

$$\chi_{FT} = \frac{\ln[P(\Delta X)/P(-\Delta X)]}{\Delta X} \quad (1)$$

where $\Delta X = X(t + \Delta t) - X(t)$ (Fig. 5b), and $P(\Delta X)$ represents the probability distribution of $\Delta X$ (Hasegawa et al. 2019; Hayashi 2018; Hayashi et al. 2018a; Hayashi et al. 2018b). $\chi_{FT}$ is introduced based on the fluctuation theorem of non-equilibrium statistical physics (Ciliberto et al. 2010; Evans et al. 1993; Seifert 2012). The theoretical background of Eq. (1) is explained in a previous reference (Hayashi et al. 2018b).

When $P(\Delta X)$ is fitted by a Gaussian function:

$$P(\Delta X) = \exp(-(\Delta X - b)^2/2a)/(2\pi a)^{0.5} \quad (2)$$

where the fitting parameters *a* and *b* correspond to the variance and mean of the distribution, respectively, $\chi_{FT}$ is practically calculated by using the quantity $\chi$:

$$\chi = 2b/a \quad (3)$$

In Fig. 7a, $\chi$ is calculated for each $P(\Delta X)$ for various intervals $\Delta t$ from 10 to 100 ms. The converged value ($\chi^*$) is related to the drag force (*F*) acting on a cargo. Previous experiments (Hasegawa et al. 2019; Hayashi et al. 2018b) suggest that

$$F \propto \chi^* \quad (4)$$

The $\chi$ value for the transport of 40 SVPs in WT *C. elegans* (Fig. 7a, left) was compared with that of 40 SVPs in *arl-8*-deletion mutant *C. elegans* (Fig. 7a, right) (Hayashi et al. 2018a). When Eq. (4) is valid, the groups represented by the different colours in Fig. 7a are regarded as force producing units (FPUs). The experimental results show four FPUs for WT *C. elegans* as compared with three FPUs for mutant *C. elegans* (Fig. 7a). Note that when 1 FPU is considered to be a dimer of UNC-104, $\chi^*\sim 0.05$ nm$^{-1}$ for 1FPU corresponds to ~ 5 pN estimated from the stall force values of UNC-104 dimer (Tomishige et al. 2002). Because the small difference of $\chi^*$ for each FPU (*e.g.* $\chi^*\sim 0.05$ nm$^{-1}$ for 1FPU, $\chi^*\sim 0.1$ nm$^{-1}$ for 2 FPUs and $\chi^*\sim 0.2$ nm$^{-1}$ for 3 FPUs) observed between WT and mutant *C. elegans* indicated that *arl-8* deletion did not affect the force generation of UNC-104, the decreased mean value of $\chi^*$ in the mutant (Fig. 7b) was considered as a consequence of the decreased number of FPUs transporting an SVP.

*Number of motors*

When 1 FPU is considered to be a dimer of UNC-104, the number of FPUs represents the number of UNC-104 dimers. Comparison of FPUs between WT and *arl-8*-deletion mutant *C. elegans* (Fig. 8a) (Hayashi et al. 2018a) indicated a decrease in the proportion of 3 FPUs, whereas that of 1 FPU increased in the case of the mutant, with a 20% decrease in the mean number of FPUs between WT and mutant (Fig. 8b). This suggested that the absence of ARL-8 decreased the number of active UNC-104 dimers capable of SVP transport.

*Velocity*

Comparison of the velocity during a CVS between WT and *arl-8*-deletion mutant *C. elegans* (*n* = 40 each) (Fig. 9) (Hayashi et al. 2018a) showed a slight reduction in the mutant relative to WT. Because both the force generated by each FPU (*e.g.* $\chi^*\sim 0.05$ nm$^{-1}$ for 1FPU, $\chi^*\sim 0.1$ nm$^{-1}$ for 2 FPUs and $\chi^*\sim 0.2$ nm$^{-1}$ for 3 FPUs shown in Fig. 7a), and the size of the SVPs being transported (Fig. 10) changed only slightly, the difference in velocity was assumed to result from the difference in the number of UNC-104 dimers transporting individual SVPs (Fig. 8). Note that velocity can depend on the number of the motors based on the model described in Fig. 6b.

*SVP fluorescence intensity*

The fluorescence intensity (FI) of SVPs is different for each SVP, representing different sized of the SVPs (Fig. 10a). The distribution of FI compared between WT and *arl-8*-deletion mutant

*C. elegans* (*n* = 40 each) (Hayashi et al. 2018a) suggests that SVP size was unchanged by *arl-8* deletion (Fig. 10b).

*Run-length and -time*

The decreased number of UNC-104 dimers transporting SVPs in the *arl-8*-deletion mutant *C. elegans* (Fig.8) is linked to a decrease in the run length and time (duration) of SVPs defined as the persistence distance and time over which an SVP continues to move without stopping (Fig. 11a). Indeed, run length increases according to the number of kinesins (Furuta et al. 2013). The run length (Fig. 11b and 11d) and time (Fig. 11c and 11e) of the mutant *C. elegans* was shorter than that of the WT.

*Pause duration*

Comparison of pause duration, representing the time interval from SVP detachment from a microtubule until its subsequent attachment to another microtubule (Fig. 11a), between WT and *arl-8*-deletion mutant *C. elegans* (Niwa et al. 2016) was 2-fold longer for the mutant relative to WT (Fig. 12). Here, pauses in SVP movement are interpreted as detachment events from a microtubule (Fig. 11a), because axon narrowness is supposed to inhibit the diffusion of SVP following detachment. Because the probability of this attachment depends upon the number of UNC-104 dimers attached to an SVP, the pause duration should be affected by the molecular number of UNC-104, as well as the run length and time (Fig. 11b-11e).

*Anterograde current*

Anterograde current is defined as the frequency of UNC-104-mediated SVP anterograde transport (Niwa et al. 2016) and is 50% diminished in the mutant relative to WT (Fig. 13).

**Measurement of physical parameters associated with synapse construction**

Here, we summarize the results of physical measurements associated with synapse constructions in DA9 motor neurons in *C. elegans* (Niwa et al. 2016) (Fig. 14). It is a future issue to explain the results from the physical parameters of UNC-104-mediated SVP transport via an appropriate physical model.

*Distance from the cell body to synaptic regions*

Fluorescence micrographs of the synaptic regions of DN9 motor neurons show that synapse location differs between WT and *arl-8*-deletion mutant *C. elegans* (Fig. 14a). Moreover, the distance from the cell body to the synaptic region (measured for this review based on a previous reference (Niwa et al. 2016)) is shorter in the mutant than in the WT (Fig. 14b). Additionally, it was reported that the synapses in the *arl-8*-deletion mutant worms localize in dendrites, as well as in the axons of DA9 motor neurons (Niwa et al. 2016).

*Distance between synaptic puncta*

Noting that the distance between synaptic puncta (Fig. 14a) characterizes synapse construction (Niwa et al. 2016), measurement of distance between synaptic puncta for neurons in WT and *arl-8*-deletion mutant *C. elagans* revealed a shorter distance in the mutant worms relative to that in WT worms.

**Discussion and perspective**

This review summarizes the physical parameters associated with UNC-104 transport of SVP in DN9 motor neurons of *C. elegans* (Fig. 2a). We focused on the mechanisms by which UNC-104 autoinhibition is released via ARL-8 (Niwa et al. 2016) (Fig. 3). These parameters measured in the previous studies (Hayashi et al. 2018a; Niwa et al. 2016) revealed the physical aspects related to SVP transport, including force (Fig. 7), the number of UNC-104 dimers carrying an SVP cargo (Fig. 8) and velocity (Fig. 9). We compared the parameters between WT and a deletion mutant of the UNC-104 regulatory gene *arl-8*. The results suggested that SVP-transport ability is weakened in the absence of ARL-8 (Fig. 7-13).

Additionally, we compared the physical parameters characterizing synapse construction in DA9 motor neurons between WT and mutant *C. elegans* (Fig. 14), with changes in these quantities indicating that UNC-104 autoinhibition is related to synaptic localization in these neurons (Niwa et al. 2016). Future construction of a physical model capable of quantitatively explaining these changes in synapse construction (Fig. 14) will be based on the physical parameters associated with UNC-104 transport of SVP (Fig. 7-13). We hope that this review provides a useful guide for the development of this model.

Furthermore, this review addresses an issue reported at the Asian Biophysics Association Symposium 2018, specifically that defective kinesin autoinhibition is related to hereditary spastic paraplegia, and that this affects certain physical parameters associated with SVP transport; therefore, this review provided information relevant to human disease. The current findings suggest that the physical parameters associated with SVP transport are useful for understanding the molecular basis of neuronal diseases related to defective motor proteins.

**Acknowledgement**

We thank the participants of the Asian Biophysics Association (ABA) Symposium 2018 for comments on the study. This work was supported by JST PRESTO (grant number JPMJPR1877), AMED PRIME (grant number JP18gm5810009), and JSPS KAKENHI (grant number 17H03659) to K. H., as well as JSPS KAKENHI (grant number 17H05010) to S. N.

**Conflicts of interest**

We declare that there are no conflicts of interest.

**Ethical approval**

All the animal experiments were conducted in compliance with protocols which was approved by the Institutional Animal Care and Use Committee, Tohoku University.

**Author contributions**

K.H. and S.N. wrote the paper with the help of S.M.. S.M. and M.G.M. performed the experiments and the data analysis (Fig. 11 and Fig. 14).

**Competing interests:** The authors declare no competing interests.

**Correspondence:** Correspondence and material requests should be addressed to K.H.

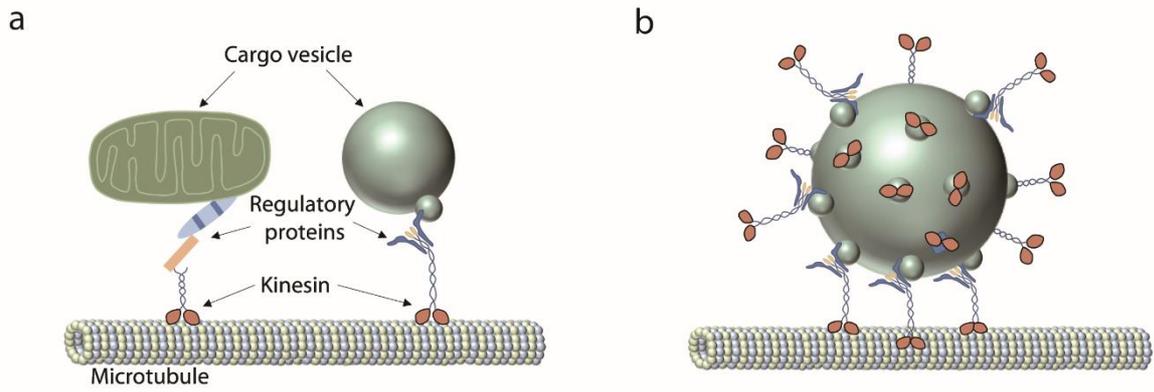

**Fig. 1** Schematics of *in vivo* cargo transport. (a) Examples of regulatory proteins that control the binding of the cargo and motor. (b) A cargo transported by multiple motors. In the schematics, three motors carry a single cargo.

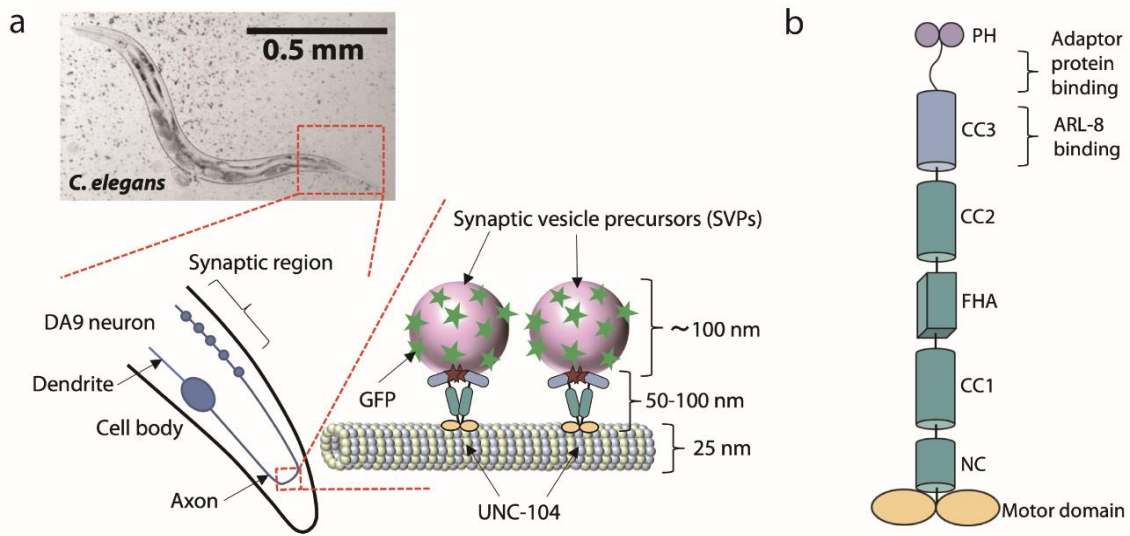

**Fig. 2** Schematics of SVP transport by UNC-104 in the axon of DA9 motor neuron of *C. elegans*. (a) For the fluorescence observation of SVPs (Fig. 5a), GFPs are attached to each SVP, allowing fluorescence-based tracking (Niwa et al. 2016). The micrograph of a *C. elegans* worm was acquired using the experimental setup described in Fig. 4a. (b) Structure of UNC-104. NC, CC, FHA and PH represent neck coiled-coil domain, coiled-coil domain, fork head associated domain and pleckstrin homology domain, respectively. PH is known to be PIP2 (phosphatidylinositol 4,5- bisphosphate) binding domain.

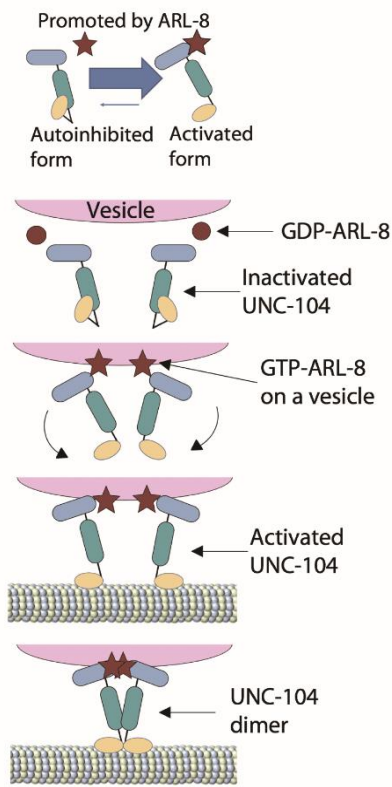 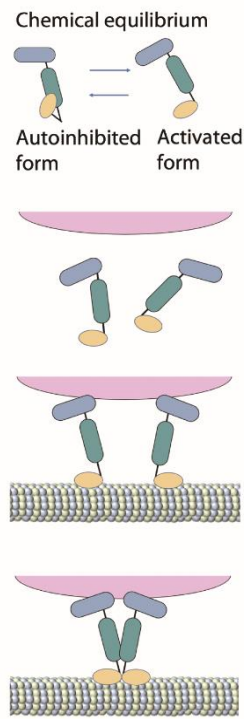

**Fig. 3** Mechanism on cargo binding of UNC-104 with (left) and without (right) ARL-8 proposed in Ref. (Niwa et al. 2016). (Left) UNC-104 is inactivated by the autoinhibition. Then, guanosine triphosphate (GTP)-ARL-8 bound to an SVP releases the autoinhibition. Note that guanosine diphosphate (GDP)-ARL-8 does not bind to SVPs. The activation of UNC-104 induces the SVP and microtubule binding. Finally, the SVP binding induces the dimerization of UNC-104 (Klopfenstein et al. 2002). (Right) Some UNC-104 monomers are in the active state without ARL-8 as a result of the chemical equilibrium between autoinhibited and activated forms of UNC-104. These UNC-104 monomers in the active state can bind to an SVP.

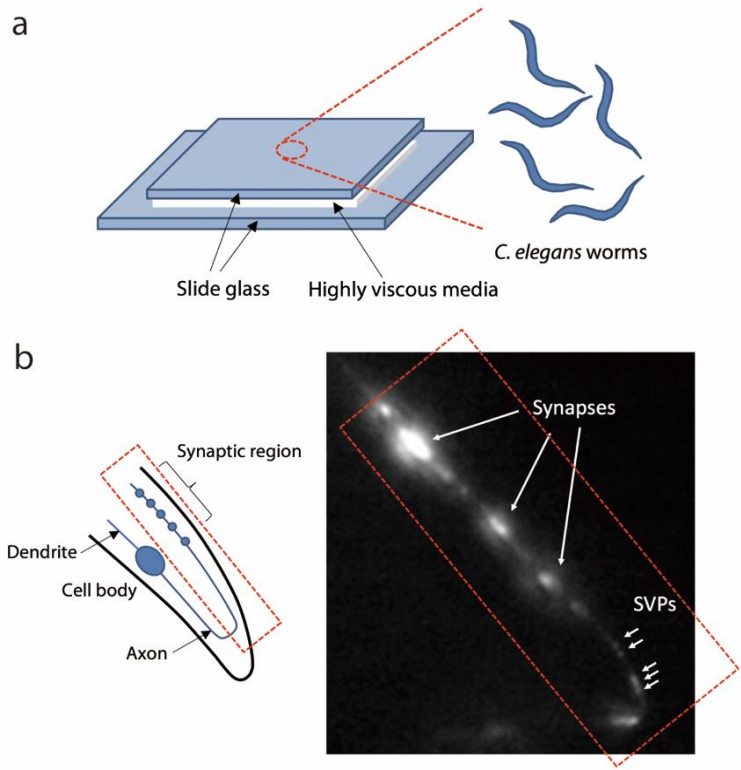

**Fig. 4** Fluorescence observation of SVPs (Hayashi et al. 2018a; Niwa et al. 2016). (a) Schematics of the cell chamber used for the fluorescence microscopy. *C. elegans* worms in highly viscous media are inserted between slide glasses. (b) Schematics (left) and fluorescence micrograph (right) of DA9 motor neuron of a *C. elegans* worm. Because many SVPs are accumulated at synapses of DA9 motor neuron where SVPs are unloaded from UNC-104, the location of synapses can be identified as strong bright spots lying around the axonal terminal regions while SVPs are identified as weak spots moving along the axon.

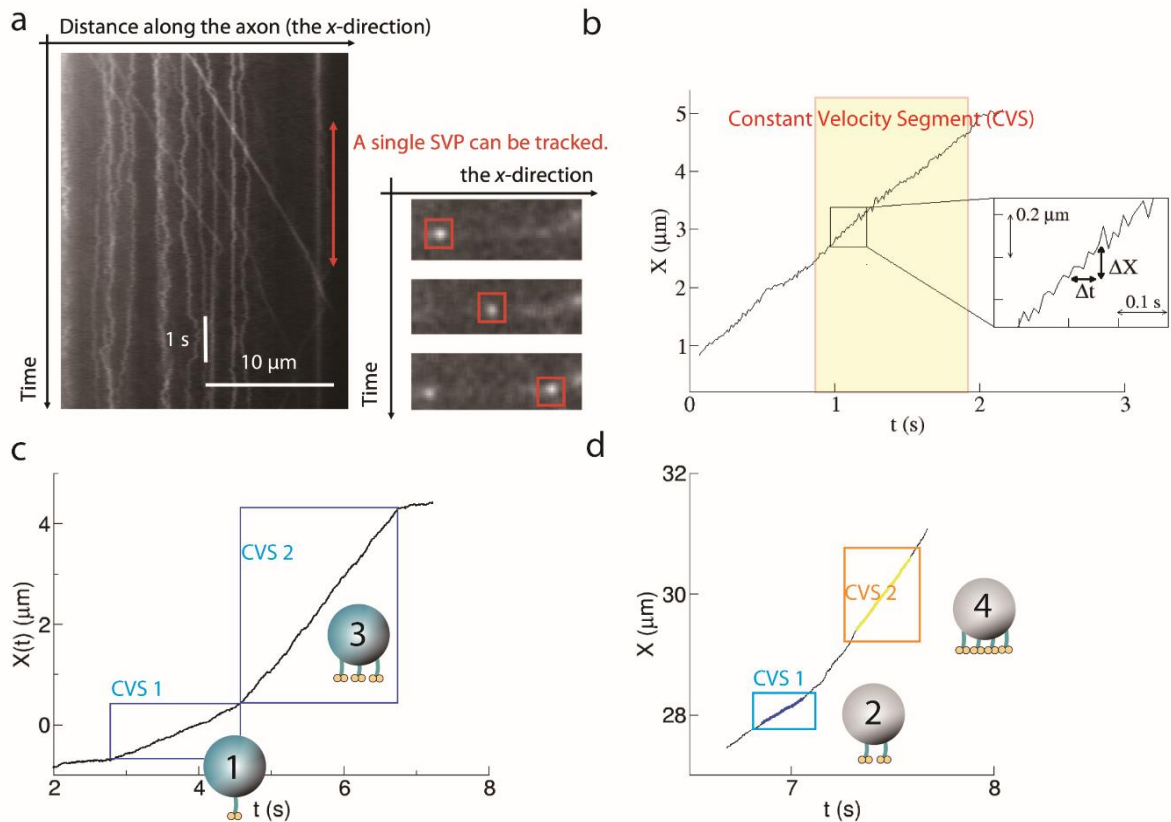

**Fig. 5** Center positions of cargo vesicles transported by motors. (a) Kymograph of SVP transport, including a time interval in which a single SVP can be tracked though SVPs are crowded in the axon. (b) Example time course of the center position of an SVP in the DA9 neuron of a *C. elegans* worm (Hayashi et al. 2018a). There is a constant velocity segment (CVS) in which an SVP moves at a constant velocity (the yellow part). Long time courses have several CVSs, and sharp velocity changes are often observed in long time courses of *in vivo* cargo transport (c-d). (c) Example time course of the center position of an endosome in the dorsal root ganglion (DRG) neuron of a mouse (Hayashi et al. 2018b). (d) Example time course of the center position of a melanosome in the melanophore of a zebrafish (Hasegawa et al. 2019).

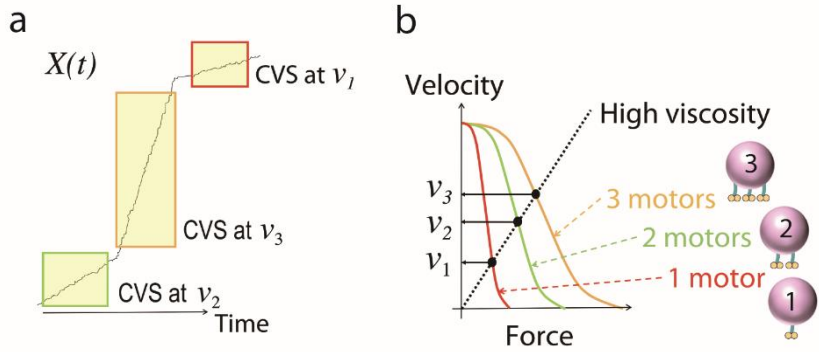

**Fig. 6** Possible explanation on the velocity change in the time course of a cargo (Fig. 5). Our assumed model is presented. (a) Schematics of velocity change in *in vivo* cargo transport. A long time course has several constant velocity segments (CVSs). (b) Schematics of the *in vivo* force-velocity relation of kinesin (Hayashi et al. 2018b). The dotted line represents the Stokes relation ($F=\Gamma v$) in the case that $\Gamma$ is large where $F$, $v$, and $\Gamma$ are force, velocity and friction coefficient of a cargo, respectively. When the number of motors carrying the cargo is changed, the velocity can be changed in this model.

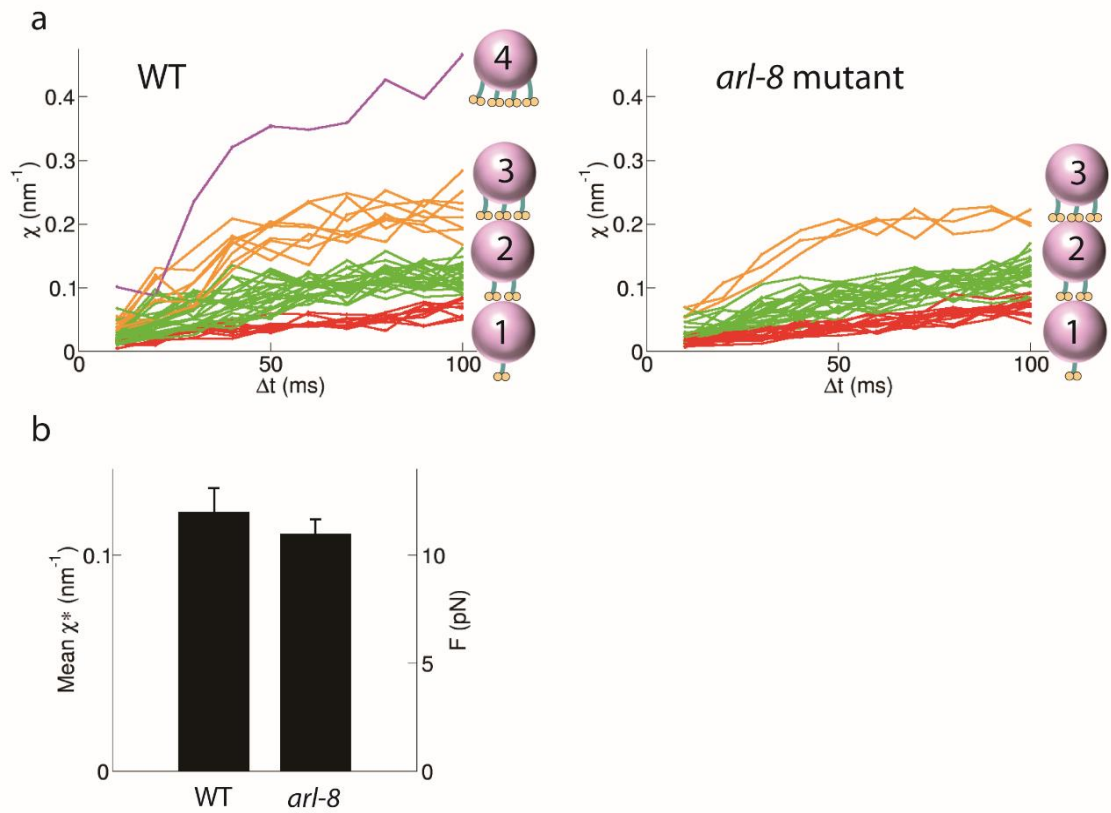

**Fig. 7** Force quantified by using the fluctuation unit ($\chi$) (Eq. (3)). (a) $\chi$ as a function of $\Delta t$ for WT *C. elegans* (left) and for the *arl-8*-deletion mutant *C. elegans* (right) (Hayashi et al. 2018a). 40 SVPs were investigated for each case. $\chi$ converges to the constant value $\chi^*$ as $\Delta t$ becomes large. (b) The mean value of $\chi^*$ is compared between WT and *arl-8*-deletion mutant *C. elegans*. In the right *y*-axis, the approximate force value is shown as a reference value, noting that $\chi^*$ is converted to the force using the stall force value of UNC-104 dimer obtained in the single-molecule experiment (Tomishige et al. 2002). The error-bars represent the standard error (n=40 for each).

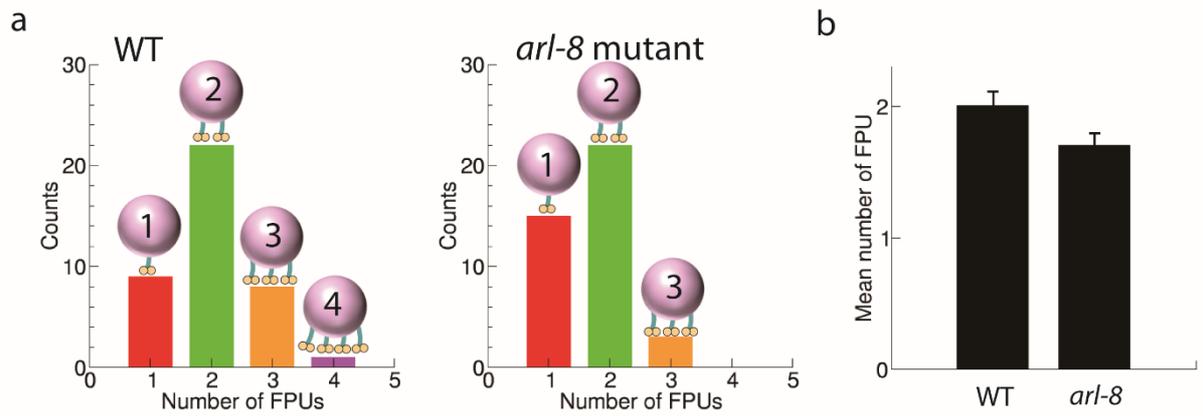

**Fig. 8** Population of force producing units (FPUs) (Hayashi et al. 2018a). (a) From the discrete behaviour of $\chi$ in Fig. 7a, the population of each FPU was calculated. The population was investigated for WT *C. elegans* (n=40) (left) and for the *arl-8*-deletion mutant *C. elegans* (n=40) (right). (b) The mean number of FPUs is compared between WT (left) and *arl-8*-deletion mutant *C. elegans*. In both cases, about two motors carry a cargo together on average. The error-bars represent the standard error (n=40 for each).

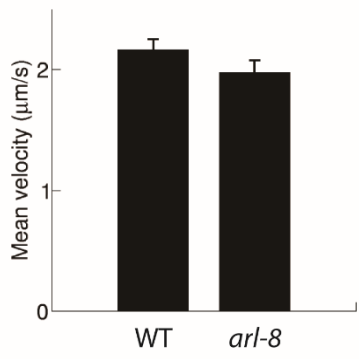

**Fig. 9** Mean velocity at constant velocity segments (CVSs) (Hayashi et al. 2018a). The mean value for velocity is compared between WT (n=40) (left) and *arl-8*-deletion mutant *C. elegans* (n=40) (right). The error bars represent the standard error (n=40 for each).

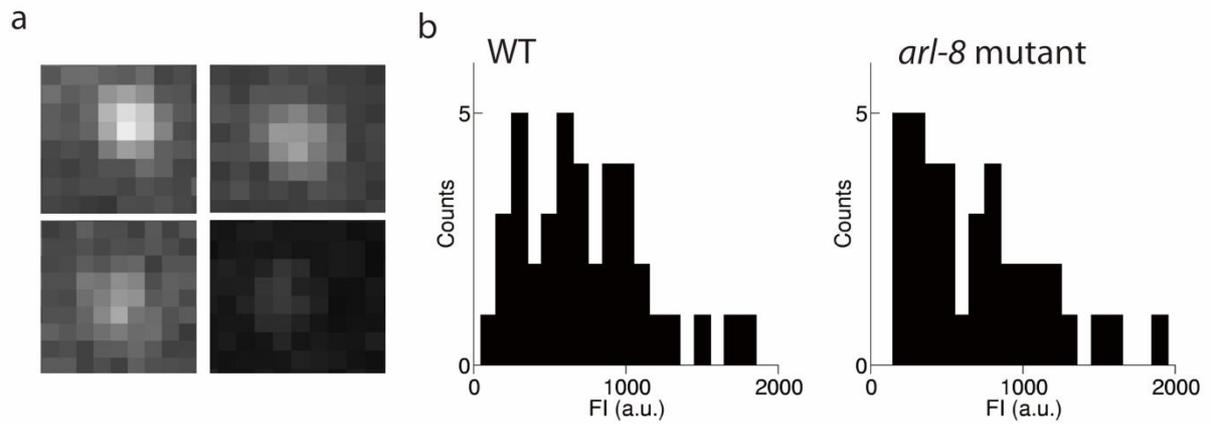

**Fig. 10** Fluorescence intensity (FI) of SVPs (Hayashi et al. 2018a). (a) The example fluorescence micrographs of SVPs. $(FI)^{1/2}$ is proportional to the size of an SVP. The size of each SVP is different. (b) The distribution of FI for WT *C. elegans* (n=40) (left) and for the *arl-8*-deletion mutant *C. elegans* (n=40) (right). Although the data is noisy, the ranges of FI values do not show big difference between WT and mutant.

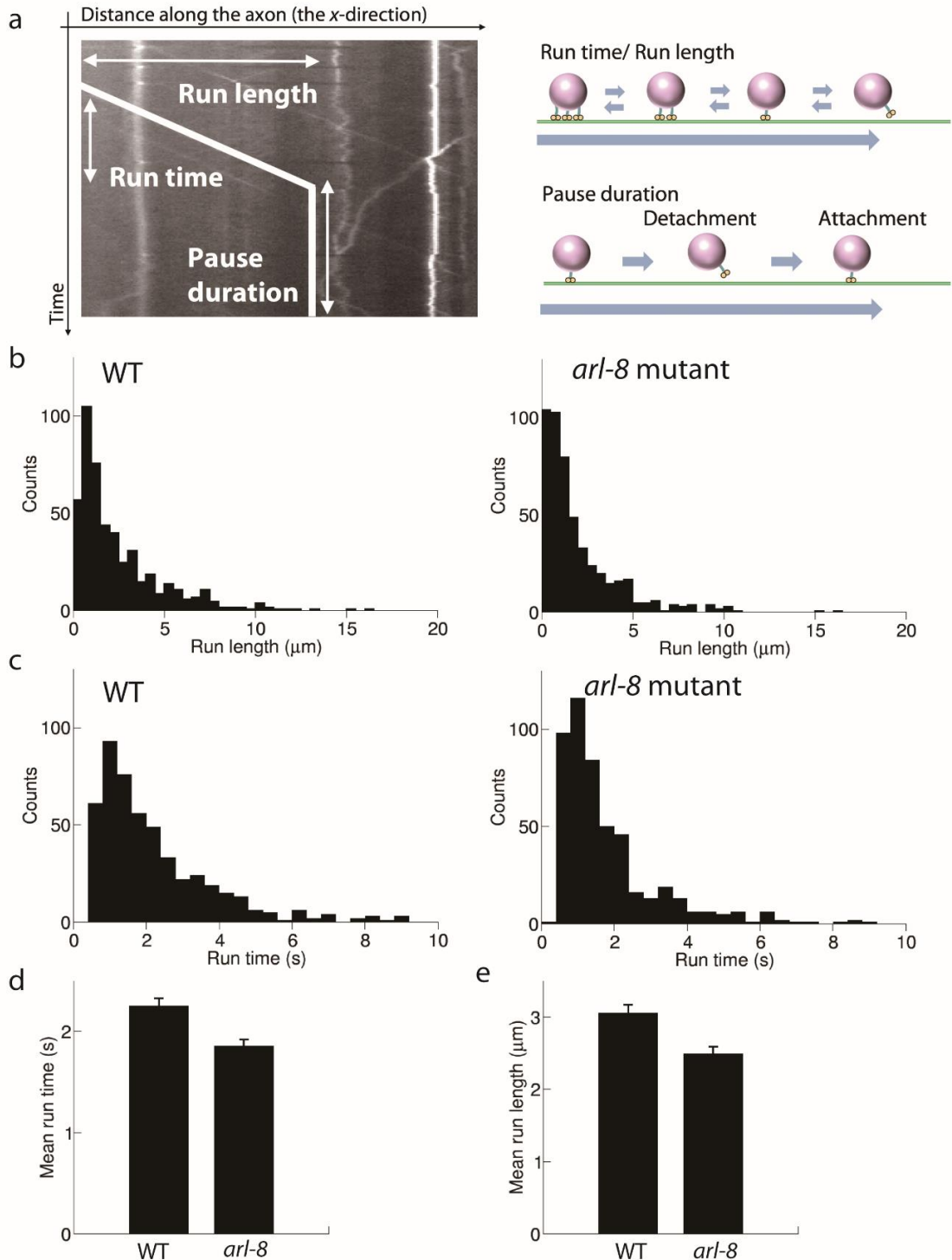

**Fig. 11** Run length and run time of SVPs. (a) Definition (left) and schematic explanation (right) of run length and run time. The distributions of run length (b) and those of run time (c) are investigated for WT *C. elegans* (n=400 from 35 different worms (18 different experiments)) and the *arl-8*-deletion mutant *C. elegans* (n=400 from 28 different worms (7 different experiments)), respectively. Mean run length (d) and mean run time (e) are compared between WT and *arl-8*-deletion mutant *C. elegans*. The error-bars represent the standard error (n=400). Note that the mean values were obtained by averaging the 400 values shown in Fig. 11b and 11c.

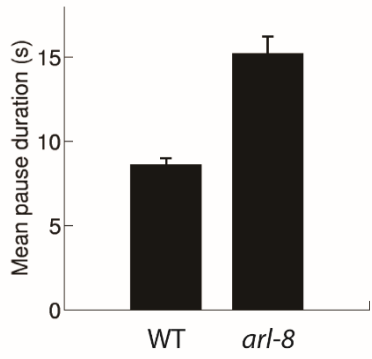

**Fig. 12** Pause duration of SVPs (Niwa et al. 2016). The definition of pause duration is described in Fig. 11a (left). Pause duration is compared between WT and *arl-8*-deletion mutant *C. elegans*. The error-bars represent the standard error (n=512 for WT, n=132 for the mutant).

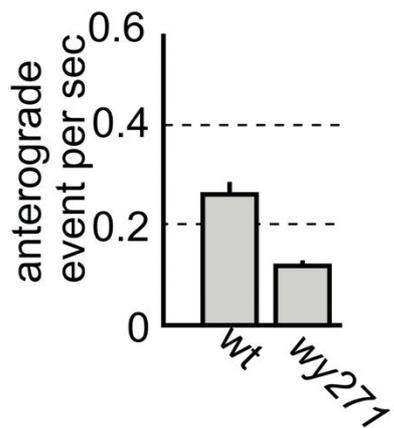

**Fig. 13** Anterograde current of SVPs (Niwa et al. 2016). Anterograde current is defined as the duration of anterograde run per second. Anterograde current is compared between WT and *arl-8*-deletion mutant *C. elegans*. The error-bars represent the standard error.

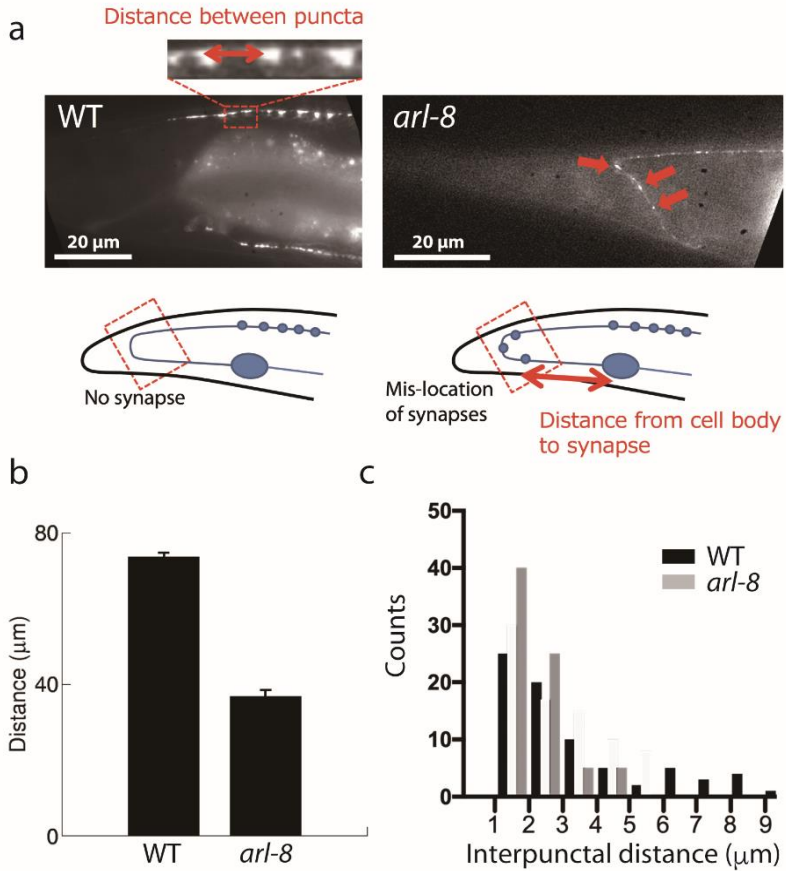

**Fig. 14** Physical measurement on synapses. (a) Fluorescence micrographs of the DA9 motor neurons in the cases of WT *C. elegans* (left) and the *arl-8*-deletion mutant *C. elegans* (right). (b) The distance from the cell body to the synaptic region is compared between WT and *arl-8*-deletion mutant *C. elegans* (n=10 for each). Note that the data was newly added for this review based on Ref. (Niwa et al. 2016). The error-bars represent the standard error. (c) The distance between synaptic puncta is compared between WT and *arl-8*-deletion mutant *C. elegans* (Niwa et al. 2016).

**References**

Chiba K et al. (2014) Quantitative analysis of APP axonal transport in neurons: role of JIP1 in enhanced APP anterograde transport Mol Biol Cell 25:3569-3580 doi:10.1091/mbc.E14-06-1111

Ciliberto S, Joubaud S, Petrosyan A (2010) Fluctuations in out-of-equilibrium systems: from theory to experiment J Stat Mech-Theory E:P12003 doi:Artn P12003

10.1088/1742-5468/2010/12/P12003

Encalada SE, Goldstein LS (2014) Biophysical challenges to axonal transport: motor-cargo deficiencies and neurodegeneration Annual review of biophysics 43:141-169 doi:10.1146/annurev-biophys-051013-022746

Evans DJ, Cohen EG, Morriss GP (1993) Probability of second law violations in shearing steady states Physical review letters 71:2401-2404 doi:10.1103/PhysRevLett.71.2401

Furuta K, Furuta A, Toyoshima YY, Amino M, Oiwa K, Kojima H (2013) Measuring collective transport by defined numbers of processive and nonprocessive kinesin motors Proceedings of the National Academy of Sciences of the United States of America 110:501-506 doi:10.1073/pnas.1201390110

Gross SP (2004) Hither and yon: a review of bi-directional microtubule-based transport Physical biology 1:R1-11 doi:10.1088/1478-3967/1/2/R01

Hall DH, Hedgecock EM (1991) Kinesin-related gene unc-104 is required for axonal transport of synaptic vesicles in C. elegans Cell 65:837-847

Hammond JW et al. (2009) Mammalian Kinesin-3 motors are dimeric in vivo and move by processive motility upon release of autoinhibition PLoS biology 7:e72 doi:10.1371/journal.pbio.1000072

Hancock WO (2016) The Kinesin-1 Chemomechanical Cycle: Stepping Toward a Consensus Biophysical journal 110:1216-1225 doi:10.1016/j.bpj.2016.02.025

Hancock WO, Howard J (1999) Kinesin's processivity results from mechanical and chemical coordination between the ATP hydrolysis cycles of the two motor domains Proceedings of the National Academy of Sciences of the United States of America 96:13147-13152

Hasegawa S, Sagawa T, Ikeda K, Okada Y, Hayashi K (2019) Investigation of multiple-dynein transport of melanosomes by non-invasive force measurement using fluctuation unit *c*. Scientific reports 9:5099 doi:10.1038/s41598-019-41458-w

Hayashi K (2018) Application of the fluctuation theorem to motor proteins: from F1-ATPase to axonal cargo transport by kinesin and dynein Biophysical reviews 10:1311-1321 doi:10.1007/s12551-018-0440-5

Hayashi K, Hasegawa S, Sagawa T, Tasaki S, Niwa S (2018a) Non-invasive force measurement reveals the number of active kinesins on a synaptic vesicle precursor in axonal transport regulated by ARL-8 Phys Chem Chem Phys 20:3403-3410 doi:10.1039/c7cp05890j

Hayashi K, Tsuchizawa Y, Iwaki M, Okada Y (2018b) Application of the fluctuation theorem for non-invasive force measurement in living neuronal axons Mol Biol Cell:mbcE18010022 doi:10.1091/mbc.E18-01-0022

Hirokawa N, Noda Y, Tanaka Y, Niwa S (2009) Kinesin superfamily motor proteins and intracellular transport Nature reviews Molecular cell biology 10:682-696 doi:10.1038/nrm2774

Kanada R, Sasaki K (2013) Energetics of the single-headed kinesin KIF1A Physical review E, Statistical, nonlinear, and soft matter physics 88:022711 doi:10.1103/PhysRevE.88.022711

Klassen MP et al. (2010) An Arf-like small G protein, ARL-8, promotes the axonal transport of presynaptic cargoes by suppressing vesicle aggregation Neuron 66:710-723 doi:10.1016/j.neuron.2010.04.033

Klopfenstein DR, Tomishige M, Stuurman N, Vale RD (2002) Role of phosphatidylinositol(4,5)bisphosphate organization in membrane transport by the Unc104 kinesin motor Cell 109:347-358

Klopfenstein DR, Vale RD (2004) The lipid binding pleckstrin homology domain in UNC-104 kinesin is necessary for synaptic vesicle transport in Caenorhabditis elegans Mol Biol Cell 15:3729-3739 doi:10.1091/mbc.e04-04-0326

Muller MJ, Klumpp S, Lipowsky R (2008) Tug-of-war as a cooperative mechanism for bidirectional cargo transport by molecular motors Proceedings of the National Academy of Sciences of the United States of America 105:4609-4614 doi:10.1073/pnas.0706825105

Nishiyama M, Higuchi H, Yanagida T (2002) Chemomechanical coupling of the forward and backward steps of single kinesin molecules Nature cell biology 4:790-797 doi:10.1038/ncb857

Niwa S (2017) Immobilization of Caenorhabditis elegans to Analyze Intracellular Transport in Neurons Journal of visualized experiments : JoVE doi:10.3791/56690

Niwa S, Lipton DM, Morikawa M, Zhao C, Hirokawa N, Lu H, Shen K (2016) Autoinhibition of a Neuronal Kinesin UNC-104/KIF1A Regulates the Size and Density of Synapses Cell reports 16:2129-2141 doi:10.1016/j.celrep.2016.07.043

Niwa S, Tanaka Y, Hirokawa N (2008) KIF1Bbeta- and KIF1A-mediated axonal transport of presynaptic regulator Rab3 occurs in a GTP-dependent manner through DENN/MADD Nature cell biology 10:1269-1279 doi:10.1038/ncb1785

Okada Y, Higuchi H, Hirokawa N (2003) Processivity of the single-headed kinesin KIF1A through biased binding to tubulin Nature 424:574-577 doi:10.1038/nature01804

Okada Y, Yamazaki H, Sekine-Aizawa Y, Hirokawa N (1995) The neuron-specific kinesin superfamily protein KIF1A is a unique monomeric motor for anterograde axonal transport of synaptic vesicle precursors Cell 81:769-780

Otsuka AJ et al. (1991) The C. elegans unc-104 gene encodes a putative kinesin heavy chain-like protein Neuron 6:113-122

Peskin CS, Oster G (1995) Coordinated hydrolysis explains the mechanical behavior of kinesin Biophysical journal 68:202S-210S; discussion 210S-211S

Sasaki K, Kaya M, Higuchi H (2018) A Unified Walking Model for Dimeric Motor Proteins Biophysical journal 115:1981-1992 doi:10.1016/j.bpj.2018.09.032

Schnitzer MJ, Visscher K, Block SM (2000) Force production by single kinesin motors Nature cell biology 2:718-723 doi:10.1038/35036345

Seifert U (2012) Stochastic thermodynamics, fluctuation theorems and molecular machines Reports on progress in physics Physical Society 75:126001 doi:10.1088/0034-4885/75/12/126001

Serra-Marques A et al. (2019) Kinesins 1 and 3 cooerate on the same vesicle to transport exocytotic carriers Biophysical journal 116:309a doi:10.1016/j.bpj.2018.11.1675

Tomishige M, Klopfenstein DR, Vale RD (2002) Conversion of Unc104/KIF1A kinesin into a processive motor after dimerization Science 297:2263-2267 doi:10.1126/science.1073386

Vale RD (2003) The molecular motor toolbox for intracellular transport Cell 112:467-480 doi:10.1016/S0092-8674(03)00111-9

Vale RD, Funatsu T, Pierce DW, Romberg L, Harada Y, Yanagida T (1996) Direct observation of single kinesin molecules moving along microtubules Nature 380:451-453 doi:10.1038/380451a0

Verhey KJ, Hammond JW (2009) Traffic control: regulation of kinesin motors Nature reviews Molecular cell biology 10:765-777 doi:10.1038/nrm2782

Visscher K, Schnitzer MJ, Block SM (1999) Single kinesin molecules studied with a molecular force clamp Nature 400:184-189 doi:10.1038/22146

Wagner OI et al. (2009) Synaptic scaffolding protein SYD-2 clusters and activates kinesin-3 UNC-104 in C. elegans Proceedings of the National Academy of Sciences of the United

States of America 106:19605-19610 doi:10.1073/pnas.0902949106

Welte MA (2004) Bidirectional transport along microtubules Current biology : CB 14:R525-537 doi:10.1016/j.cub.2004.06.045

Wu YE, Huo L, Maeder CI, Feng W, Shen K (2013) The balance between capture and dissociation of presynaptic proteins controls the spatial distribution of synapses Neuron 78:994-1011 doi:10.1016/j.neuron.2013.04.035

Zhou HM, Brust-Mascher I, Scholey JM (2001) Direct visualization of the movement of the monomeric axonal transport motor UNC-104 along neuronal processes in living Caenorhabditis elegans The Journal of neuroscience : the official journal of the Society for Neuroscience 21:3749-3755